\documentclass{elsart}
\usepackage{amssymb}
\usepackage[numbers]{natbib}
\usepackage{ifpdf}
\ifpdf
\usepackage[pdftex]{graphicx}
\usepackage[pdftex]{hyperref}
\usepackage[pdftex,hyperref]{xcolor}
\else
\usepackage[dvips]{graphicx}
\usepackage[dvips]{hyperref}
\usepackage[dvips,hyperref]{xcolor}
\fi
\hypersetup{colorlinks=false, pdfpagemode=UseOutlines, urlcolor=blue,  pdfstartview=FitH, pdfpagelayout=OneColumn}

\newcommand{\unit}[1]{\;\mathrm{#1}}
\newcommand{\hunit}[1]{\textrm{-}\mathrm{#1}}

\journal{arXiv}
\begin{document}
\begin{frontmatter}
\title{Saturated nucleate pool boiling of oxygen under magnetically-enhanced effective gravity}
\author[cit]{T.A. Corcovilos\corauthref{tc}}\ead{corcoted@caltech.edu},
\author[cit]{M.E. Turk},
\author[jpl]{D.M. Strayer},
\author[cit]{N.N. Asplund},
\author[cit]{N.-C. Yeh}
\address[cit]{Division of Physics, Math, and Astronomy, California Institute of Technology, Pasadena, CA, 91125}
\address[jpl]{Jet Propulsion Laboratory, California Institute of Technology, Pasadena, CA, 91109}
\corauth[tc]{Corresponding author.}
\begin{abstract}
We investigate the effect of enhancing gravity on saturated nucleate pool boiling of oxygen for effective gravities of $1g$, $6.0g$, and $16g$ ($g=9.8\unit{m s^{-2}}$) at a saturation pressure of $760\unit{\textrm{torr}}$ and for heat fluxes of $10\sim 3000\unit{W m^{-2}}$.  The effective gravity on the oxygen is increased by applying a magnetic body force generated by a superconducting solenoid.  We measure the heater temperature (expressed as a reduced superheat) as a function of heat flux and fit this data to a piecewise power-law/linear boiling curve.  At low heat flux ($\lesssim 400\unit{W m^{-2}}$) the superheat is proportional to the cube root of the heat flux. At higher heat fluxes, the superheat is a linear function of the heat flux.  To within statistical uncertainties, which are limited by variations among experimental runs, we find no variation of the boiling curve over our applied gravity range.
\end{abstract}
\end{frontmatter}

\section{Introduction}
Nucleate pool boiling is one of the enduring problems in fluid mechanics and thermal physics.  The large number of physical variables, many of them not directly measurable or controllable, and the difficulty in reproducing exact experimental conditions have thwarted attempts at constructing an accurate theoretical description, or even phenomenological description, of this everyday process.  In an attempt to shed more light on this difficult topic, we have chosen to study the effect of varying one particular parameter which has been inaccessible to most investigators: gravity.

Many widely-used empirical correlations of boiling heat transfer neglect gravity.  Several make no mention of gravity at all (Cooper\cite{cooper84a}, Stephan \& Ab\-del\-sal\-am\cite{stephan80}).  Those that do include the gravitational acceleration $g$ (Rohsenow\cite{rohsenow52}, Forster \& Greif\cite{forster59}), often do so only as a dimensional constant, not as a physical variable.  In these models varying the value of $g$ yields unphysical predictions, as was recognized by Dhir\cite{dhir99}, who suggests that these correlations are only valid when $g$ is held constant at its Earth value of $9.8\unit{m s^{-2}}$, regardless of the actual gravity in the experiment.  To develop adequate models which account for gravity variations, more experimental data are needed.

Experiments at high effective gravities using centrifuges\cite{merte90, ulucakli90} and parabolic-trajectory aircraft\cite{kim02} show results which vary depending on the applied heat flux. At low heat fluxes, increasing gravity increases heat transfer.  At high heat fluxes, however, increasing gravity decreases heat transfer\cite{merte90}.

Low gravity studies show an even wider range of results.  The general consensus among investigators is that the factors influencing nucleate boiling can be qualitatively divided into macroscopic and microscopic phenomena.  Macroscopic effects, such as buoyancy, bubble dynamics, and thermocapillary effects, tend to depend on gravity.  Microscopic effects at the heater surface, such as intermolecular forces, microlayer evaporation and microlayer conduction, are independent of gravity.  As gravity is reduced, the microscopic phenomena begin to dominate over the macroscopic effects and compete with one another, resulting in varying experimental results depending on the details of the experiments\cite{kim03, dimarco03, ohta03}.

In this article, we investigate the effect of enhanced gravity in our Earth-based laboratory; to do this we must apply some additional body force to the fluid.  Previous studies have used mechanical forces to create this body force\cite{merte90, ulucakli90, kim02}.  However, we chose to use the large magnetic susceptibility of oxygen to produce the additional body force by applying a relatively modest magnetic field.  By utilizing existing equipment in our laboratory, in particular a $10\hunit{cm}$ bore, $9\hunit{T}$ solenoidal superconducting magnet, we have constructed an inexpensive test apparatus for investigating a range of net effective gravities $g_\mathrm{eff}$ between microgravity and $90g$, with $g=9.8\unit{m s^{-2}}$ representing the Earth's gravitational acceleration.  Here we present data for saturated nucleate pool boiling under modestly increased gravities, between $1g$ and $16g$ of net effective gravity.
  
\section{Basic principle}
We simulate enhanced gravity by applying a body force to the test fluid that is the result of the interaction of the fluid's magnetic susceptibility with an applied magnetic field.  The magnetic acceleration $\mathbf{a}$ on a fluid with magnetic susceptibility $\chi$ in a magnetic field $\mathbf{B}$ can be expressed as
\begin{equation}
\mathbf{a} = \frac{1}{\mu_0} \frac{\chi}{\rho} \left(\mathbf{B} \cdot \nabla\right) \mathbf{B},
\end{equation}
where $\rho$ is the mass density of the fluid, $\mu_0$ is the vacuum permeability, and all quantities are in SI units.  For oxygen near the normal boiling point ($90.2\unit{K}$) the ratio $\chi/\rho$ equals $+3.02 \times 10^{-6}\unit{m^3 kg^{-1}}$ for both liquid and gas phases\cite{crc98}.  We assume any temperature dependence in $\chi/\rho$ is negligible over our temperature range of $90 - 95\unit{K}$.  To achieve $1g$ of acceleration on oxygen at the normal boiling point requires $\left| \left(\mathbf{B} \cdot \nabla\right) \mathbf{B} \right| = 4.5\unit{T^2m^{-1}}$. The magnetic force causes an acceleration that is independent of the fluid density and therefore independent of the phase of the fluid.

\begin{figure}[p]
	\begin{center}
	\includegraphics[width=\textwidth]{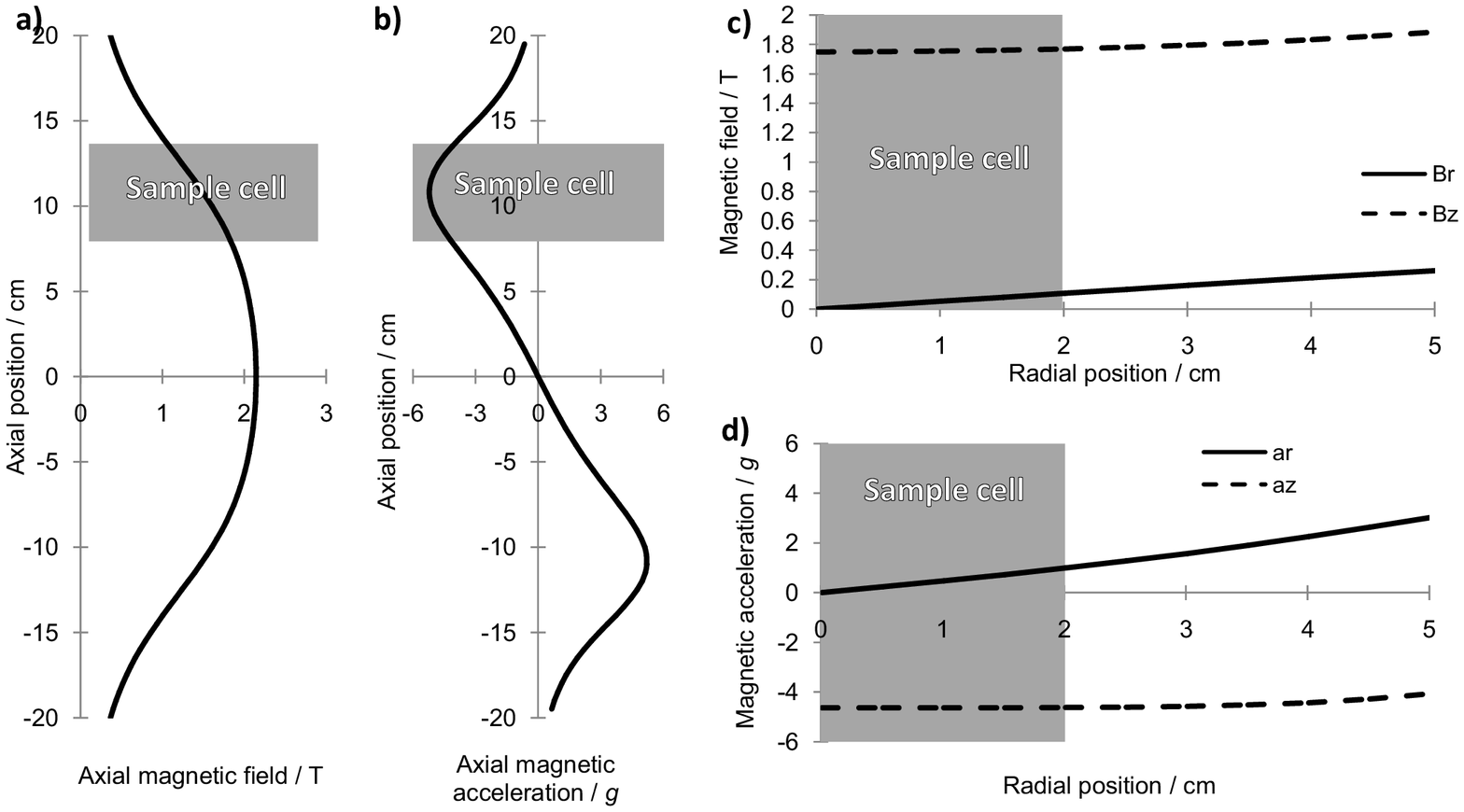}
	\caption{Plots of the magnetic field and resulting acceleration for a magnet current of $20\unit{A}$.  Panel a: On-axis magnetic field $B_z$ as a function of axial position $z$.  Panel b: On-axis magnetic acceleration $a_z$ on oxygen as a function of axial position $z$.   Panel c: axial component of magnetic field $B_z$ (dashed line) and radial component $B_r$ (solid line) evaluated at the heater surface ($z=8.8\unit{cm}$) as a function of distance from the axis $r$.  Panel d: axial component of magnetic acceleration $a_z$ (dashed line) and radial component $a_r$ (solid line)  evaluated at the heater surface ($z=8.8\unit{cm}$) as a function of distance from the axis $r$.  The additional $-1g$ of acceleration in the $-z$ direction from the Earth's gravity is not shown in these plots.  The approximate location of the sample cell ($4.4\unit{cm}$ in height, centered at $z=11.0\unit{cm}$, and $2.3\unit{cm}$ in radius) is indicated by the shaded areas.  Not shown is the size of the magnet, which extends from $-15\unit{cm}<z<+15\unit{cm}$.}
	\label{fig:force}
	\end{center}
\end{figure}

To produce our magnetic field we use a superconducting solenoid manufactured by American Magnetics, Inc. (Oak Ridge, TN, USA).  The field of the magnet is approximately that of a thin uniformly-wound finite solenoid.  In particular, the on-axis magnetic field (shown in Figure \ref{fig:force}a) is given by
\begin{equation}
B_z(z) \approx I C \left( \frac{z+a/2}{\sqrt{(z+a/2)^2+b^2}}-\frac{z-a/2}{\sqrt{(z-a/2)^2+b^2}}
 \right),
\end{equation}
with parameters $a = 25.5\unit{cm}$, $b = 7.78\unit{cm}$ and $C = 6.28\times 10^{-2}\unit{T\,A^{-1}}$ (determined by fitting to data provided by the manufacturer), $I$ being the current in the magnet measured in amperes (maximum of $84\unit{A}$ for this magnet), and the axial position $z$ measured relative to the center of the magnet.  The corresponding on-axis applied acceleration on the oxygen is plotted in Figure \ref{fig:force}b.  The off-axis fields at the heater surface ($z = 8.8\unit{cm}$), calculated using formulas in Ref. \cite{conway01}, and the resulting accelerations are plotted  in Fig. \ref{fig:force}c and \ref{fig:force}d, respectively.  We ignore the radial force in our analysis.

To obtain an approximately flat force profile over the volume of the liquid, we place the sample cell near the end of the magnet where the force obtains its maximum magnitude, with the center of the sample cell $11.0\unit{cm}$ above the magnet center, as illustrated in Fig. \ref{fig:force}.  Future studies at reduced effective gravity may be performed by placing the sample at the bottom end of the magnet, where the magnetic force counteracts the Earth's gravity.

\section{Apparatus}
\begin{figure}[p]
	\begin{center}
	\raisebox{2.8in}{{\footnotesize a)}}\includegraphics[width=2.5in]{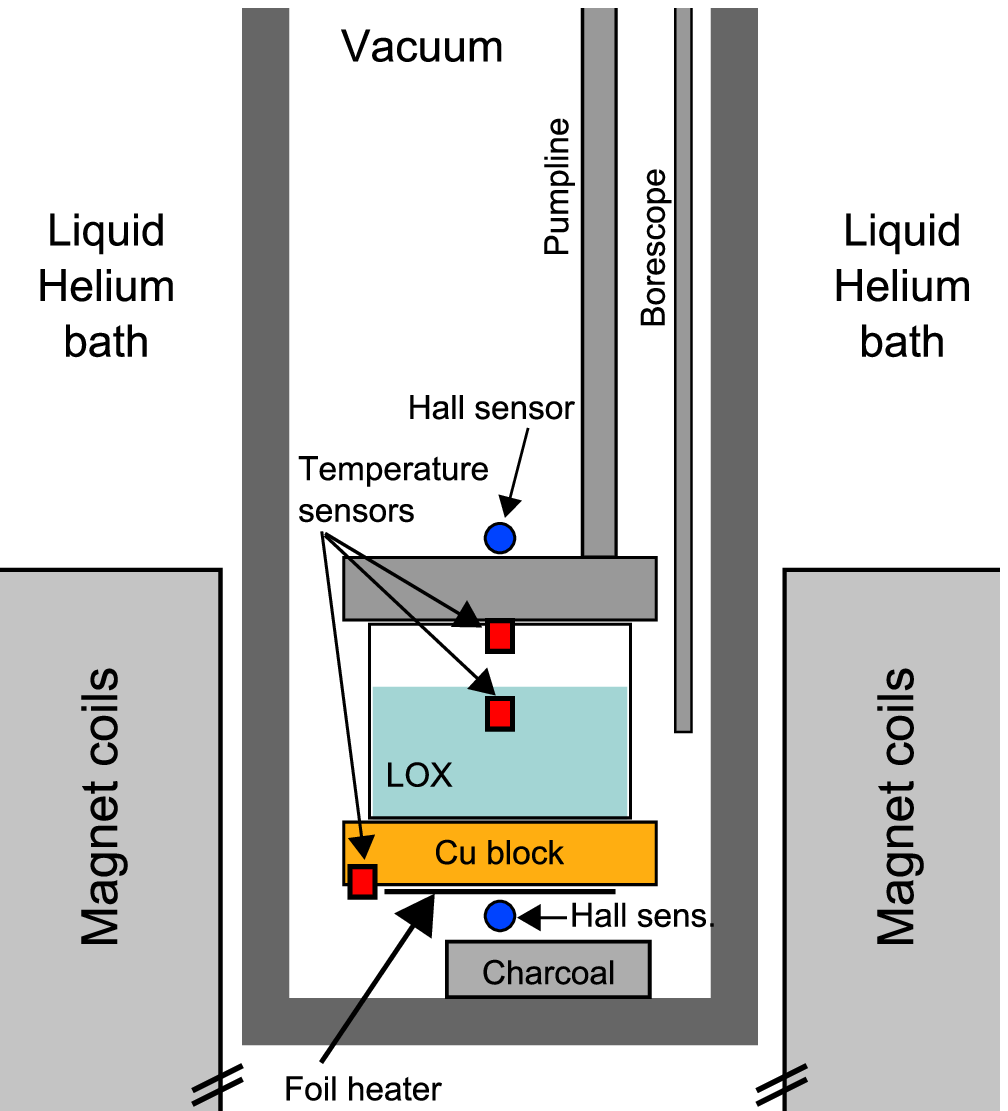}
	\raisebox{2.8in}{{\footnotesize b)}}\includegraphics[width=2.5in]{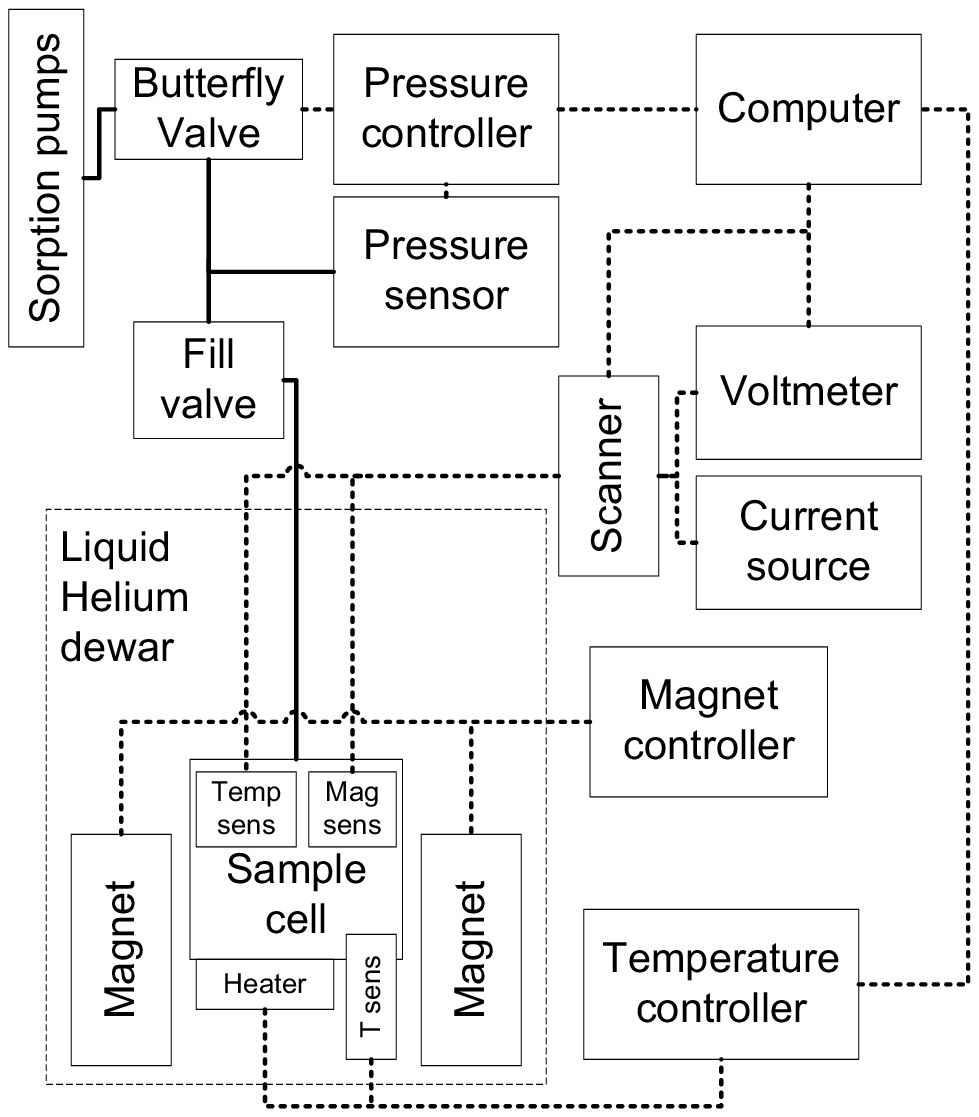}
	\caption{Panel a: a schematic of the experimental setup (not to scale).  The sample cell containing the liquid oxygen (LOX) consists of a hollow quartz cylinder, stainless steel top plate, and a copper bottom plate.  A thin foil heater is attached to the bottom of the copper plate, outside the sample cell.  Resistive temperature sensors (red boxes) are located throughout the sample cell with an additional sensor embedded in the copper plate.  Hall effect magnetic field sensors (blue circles) are located at either end of the sample cell, along the magnet axis.  The cell is located in the vacuum space of the cryogenic probe, which is itself immersed in the liquid helium used to cool the magnet. Panel b: a block diagram of controls.  The pumplines are indicated by solid lines.  Electronic connections are shown by thick dashed lines.}
	\label{fig:block}
	\end{center}
\end{figure}

A schematic of our experimental setup is shown in Figure \ref{fig:block}a.  The sample cell is a quartz cylinder ($50\unit{mm}$ O.D., $2\unit{mm}$ thickness, and $43\unit{mm}$ height, inner volume of $70\unit{cm^3}$) with a  stainless steel top plate and an oxygen-free high conductivity (OFHC) copper bottom plate (area of $16.6\unit{cm^2}$), joined together by indium gaskets held in compression with 6-32 bolts tightened to $5\unit{N\cdot m}$ of torque.  Lakeshore model CX-1050-SD Cernox temperature sensors are placed on the interior surface of the top plate, in the center of the cell, and embedded in the bottom plate.  The bottom sensor measures the heater plate temperature and is the primary temperature sensor used for analysis of the experiment.

The top and center temperature sensors are used only during sample filling.  During the boiling phase of the experiment the temperatures measured by these sensors drift depending on the liquid height and therefore are not useful for quantitative analysis.

For the heat source, a 75-$\Omega$ Minco resistive foil heater is attached to the exterior of the copper plate using Apiezon N grease and a beryllium copper clamp.  The interior surface of the copper plate is roughened using 200-grit sandpaper (to provide nucleation sites) and then cleaned in consecutive washes of detergent, acetone, and isopropanol.  Between experiments the sample cell is kept under vacuum to prevent adsorption of gas onto the heater surface which could change the boiling dynamics.  To measure the magnetic field, Toshiba THS-122 Hall-effect sensors are located on each end of the sample cell, on-axis.
A fiberoptic borescope built by Myriad Fiber (Dudley, MA, USA) is positioned adjacent to the quartz sample cell wall to provide imaging of the experiment.  The borescope image is output via a Panasonic GP-KS162HD CCD camera to a television monitor.  The temporal and spatial resolution of the camera are insufficient for quantitative measurements of bubble dynamics but serve nicely for diagnostic purposes and qualitative observations such as the level of the liquid during filling and identification of bubbles to confirm boiling.

The heater and the bottom temperature sensor are controlled by a Lakeshore LS-340 temperature controller.  The other temperature sensors and the magnetic field sensors are measured in a 4-lead configuration using Keithley 182 voltmeters and Keithley 220 current sources.

A $6\hunit{mm}$ O.D. stainless steel pumpline extends from the top of the sample cell to room-temperature gas handling equipment.  Pumping of the oxygen sample gas is performed by Varian Vacsorb sorption pumps, filled with zeolite getter and cooled by liquid nitrogen.  In consideration of the safety challenges of working with oxygen gas at potentially high pressures, we chose these pumps for their low cost and ease of use relative to fluorinated lubricant filled mechanical pumps.  An MKS 651C pressure regulator, actuating an MKS 253A butterfly value, controls the system pressure, which is measured by an MKS 626A Barotron sensor placed in front of the butterfly valve.

All of the above electronic sensing and control equipment are connected via GPIB to a PC running LabVIEW 7.0 software.  The sensors are polled sequentially with a cycle of approximately 10 seconds.  The control relationships between the various components are summarized in Figure \ref{fig:block}b.

\section{Procedure}
We prepare an experimental run by first precooling the magnet dewar with liquid helium and precooling the sample probe in a separate liquid nitrogen dewar.  Approximately 0.1 torr of helium gas is added to the vacuum space of the probe to facilitate heat exchange between the sample cell and the nitrogen bath during filling.  We fill the sample cell by introducing room-temperature oxygen gas into the sample cell via the pumpline.  The gas condenses until the cell is filled with liquid oxygen, a process which takes about 90 minutes.  The liquid level can be monitored visually using the borescope and can also be inferred by monitoring the various temperature sensors inside the cell.  After the oxygen fills the sample cell, we close the fill valve and remove the helium exchange gas from the vacuum space using a turbomechanical pump, typically pumping overnight to a pressure of $10^{-6}\unit{\textrm{torr}}$.  The probe is then moved to the liquid-helium-filled magnet dewar.  As the probe exterior cools to liquid helium temperature, a charcoal pot attached to the interior surface of the vacuum can absorbs any residual exchange gas in the vacuum space.

\begin{figure}[p]
	\centering
		\includegraphics[width=\textwidth]{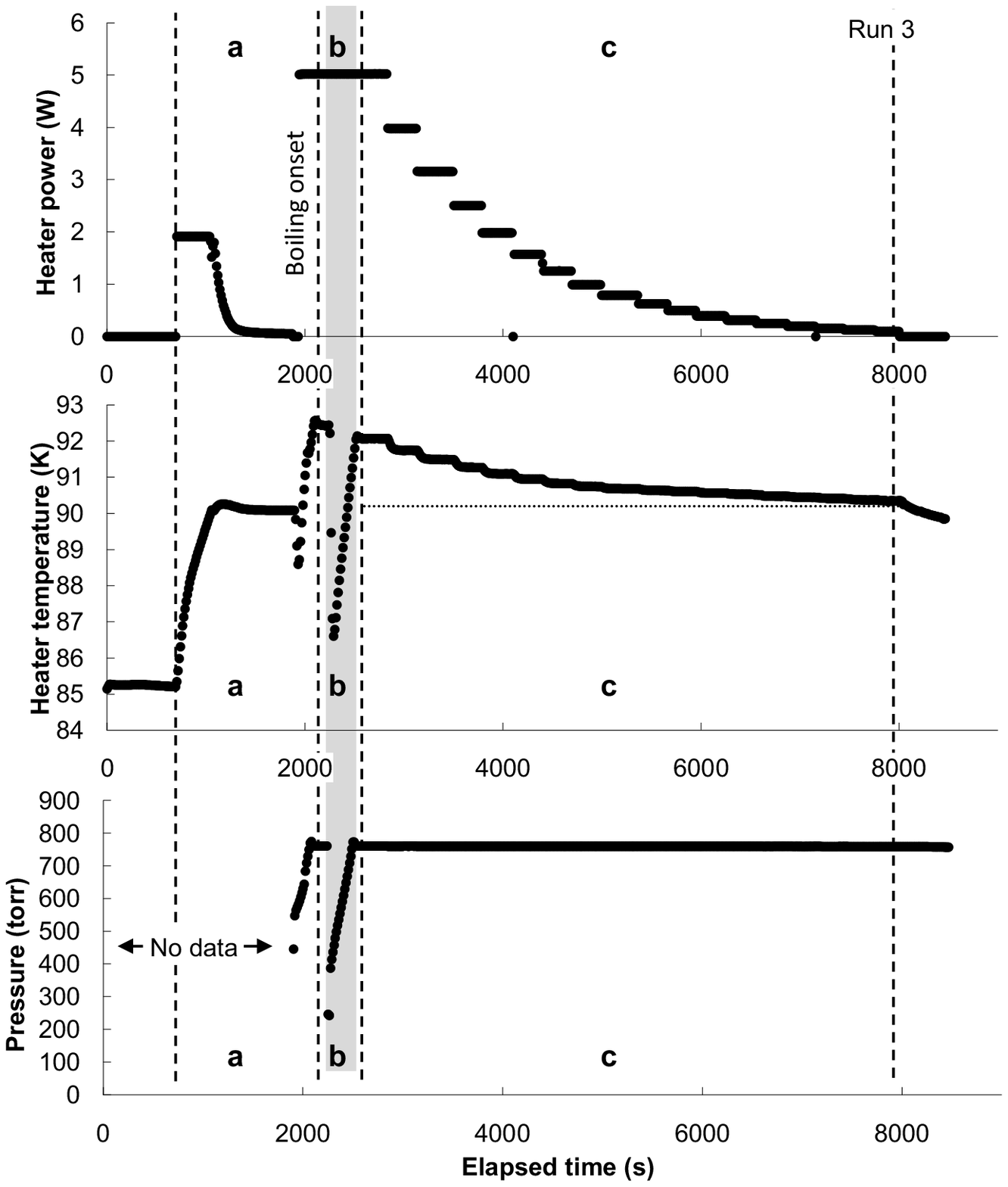}
	\caption{A typical data run at a saturation pressure of $760\unit{torr}$ and an effective gravity of $6.0g$.  The horizontal axis (common to all three frames) shows elapsed time in seconds.  The vertical axes show (from top to bottom) applied heater power, heater temperature, and system pressure.  Some time regions of interest are separated by dashed lines and labeled in the bottom frame: a) warm-up and stabilization, b) transient behavior (shaded region), where the onset of boiling causes pressure and temperature oscillations until the pressure controller stabilizes, and c) data collection.  The horizontal dotted line in the center panel indicates the saturation temperature.  There are no pressure data in region ``a'' because the valve to the sample cell is closed during this phase of the experiment.}
	\label{fig:run}
\end{figure}

We begin monitoring the temperature and magnetic field sensors at this time.  Figure \ref{fig:run} contains heater power, heater temperature, and system pressure data taken during a typical experimental run.  Details of the various stages of the experiment are described below.

To prevent the sample from freezing during the initial setup of the experiment (region ``a'' of Fig. \ref{fig:run}), we maintain the heater temperature (measured by the sensor embedded in the copper plate, see Fig. \ref{fig:block}a) at $90\unit{K}$ (just below the $1\unit{atm}$ boiling point) using a Lakeshore LS-340 temperature controller.  We record the heater power required to maintain this constant temperature with all valves closed, so that the system volume is fixed, and interpret this value as heat lost to the liquid helium bath by conduction through the support structures of the probe and the residual helium gas in the vacuum space; we subtract this quiescent heater power (typically, $\sim 50\unit{mW}$) from later heat measurements.  While the temperature of the sample probe is stabilizing, we increase the magnet current to the desired value.

To begin the experiment proper, we open the valves to the sample and set the pressure controller to the desired value ($760\unit{\textrm{torr}}$ for the data presented here).  We disable the temperature control loop and set the heater power to the desired maximum value for the run ($5.0\unit{W}$ for the data presented here, corresponding to a maximum heat flux of $3.0\unit{kW\,m^{-2}}$).  As the temperature of the sample increases, the transition from natural convection to boiling appears as a sudden drop in the heater temperature because of the increased heat transfer efficiency in the boiling state (see the dashed line at the beginning of region ``b'' in Fig. \ref{fig:run}, although see the discussion below).  The temperature then continues to increase until the system reaches some initial steady-state with approximately constant temperature and pressure.

After the system stabilizes at this initial heater power, we decrement the power by steps, typically in geometric series (region ``c'' in Fig. \ref{fig:run}).  At each new value of the power, we wait for the system to stabilize (about 5 minutes) and record the new steady-state heater temperature.  The experiment ends when the heater power decreases below the threshold to maintain a constant temperature.  This threshold heater power typically agrees within $10\%$ of the quiescent heater power measured during the setup phase of the experiment.

The time evolution of temperature and pressure during the initial warm-up (region ``b'' in Fig. \ref{fig:run}) varies between runs because of transient behavior in the pressure regulation control  loop.  A common occurrence is that as the system transitions from natural convection to boiling (characterized by a drop in heater temperature), the sudden increase in gas flow upsets the feedback loop of the pressure controller, causing large pressure and temperature oscillations for several minutes while the controller re-establishes steady-state behavior.  Some consequences of this transient behavior are discussed in the sections below.

\section{Analysis and discussion}
To characterize our data, we first rescale the heater temperature $T$ to a reduced superheat $t$ relative to the saturation temperature of the oxygen $T_b$:
\begin{equation}
t = \frac{T-T_b}{T_b}.
\end{equation}
The saturation temperature $T_b$ is calculated by measuring the system pressure on the external pumpline (see Fig. \ref{fig:block}b) and then applying the equilibrium equation of state\cite{nist12}.  
To directly measure the saturation temperature would require placing temperature sensors at the liquid-gas interface.  Because the liquid level changes throughout the experiment our fixed-position sensors are inadequate for this measurement.

The purpose of changing variables to the reduced superheat $t$ is to provide a dimensionless parameter for analysis and facilitate comparison with future measurements at varying saturation pressures.  We anticipate, based on some preliminary measurements not presented here, that
 the boiling curves at different saturation pressures will collapse onto a single curve when expressed in these units.

\begin{figure}[p]
	\centering
\raisebox{1.5in}{{\footnotesize a)}}\includegraphics[height=1.6in]{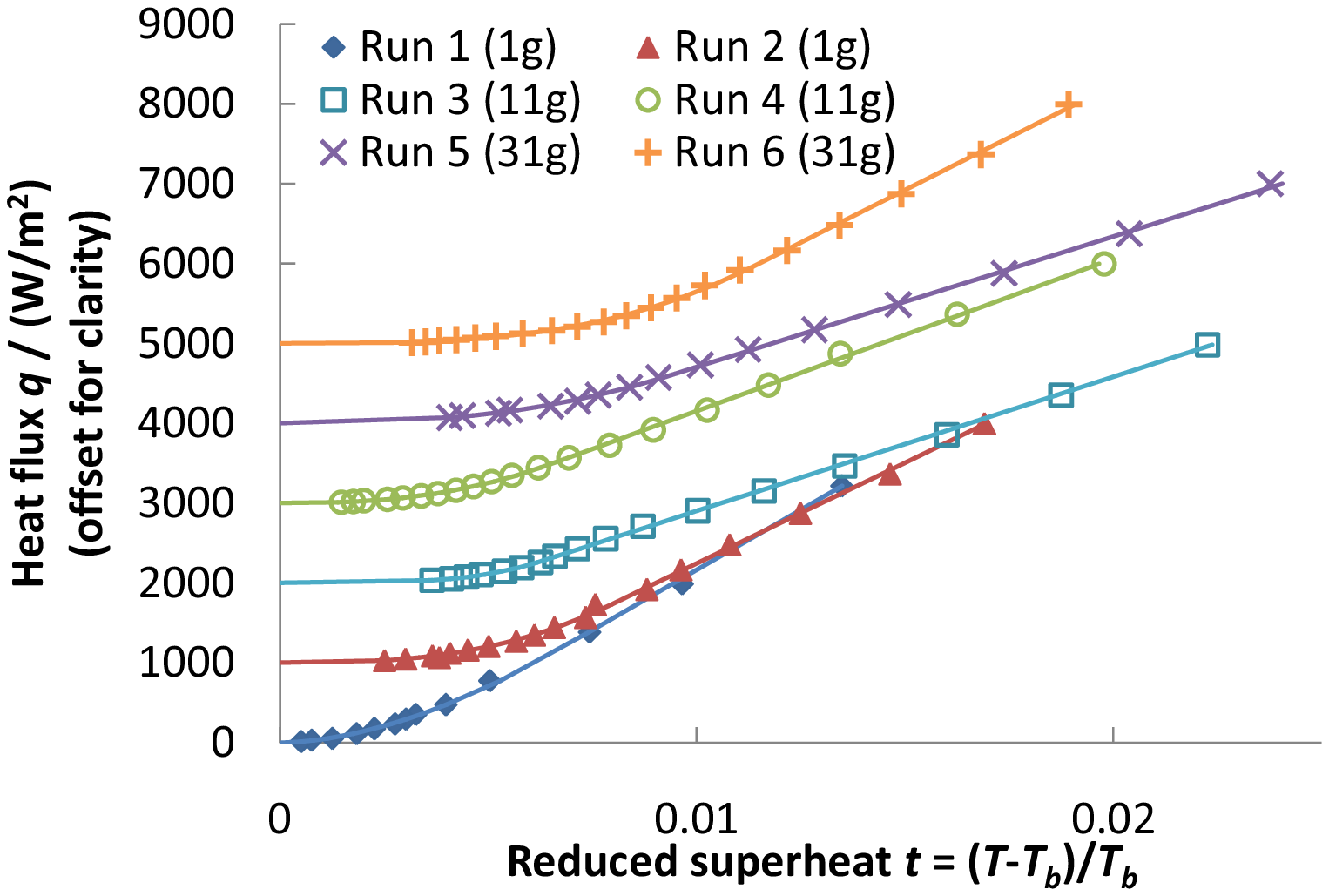}
\raisebox{1.5in}{{\footnotesize b)}}\includegraphics[height=1.6in]{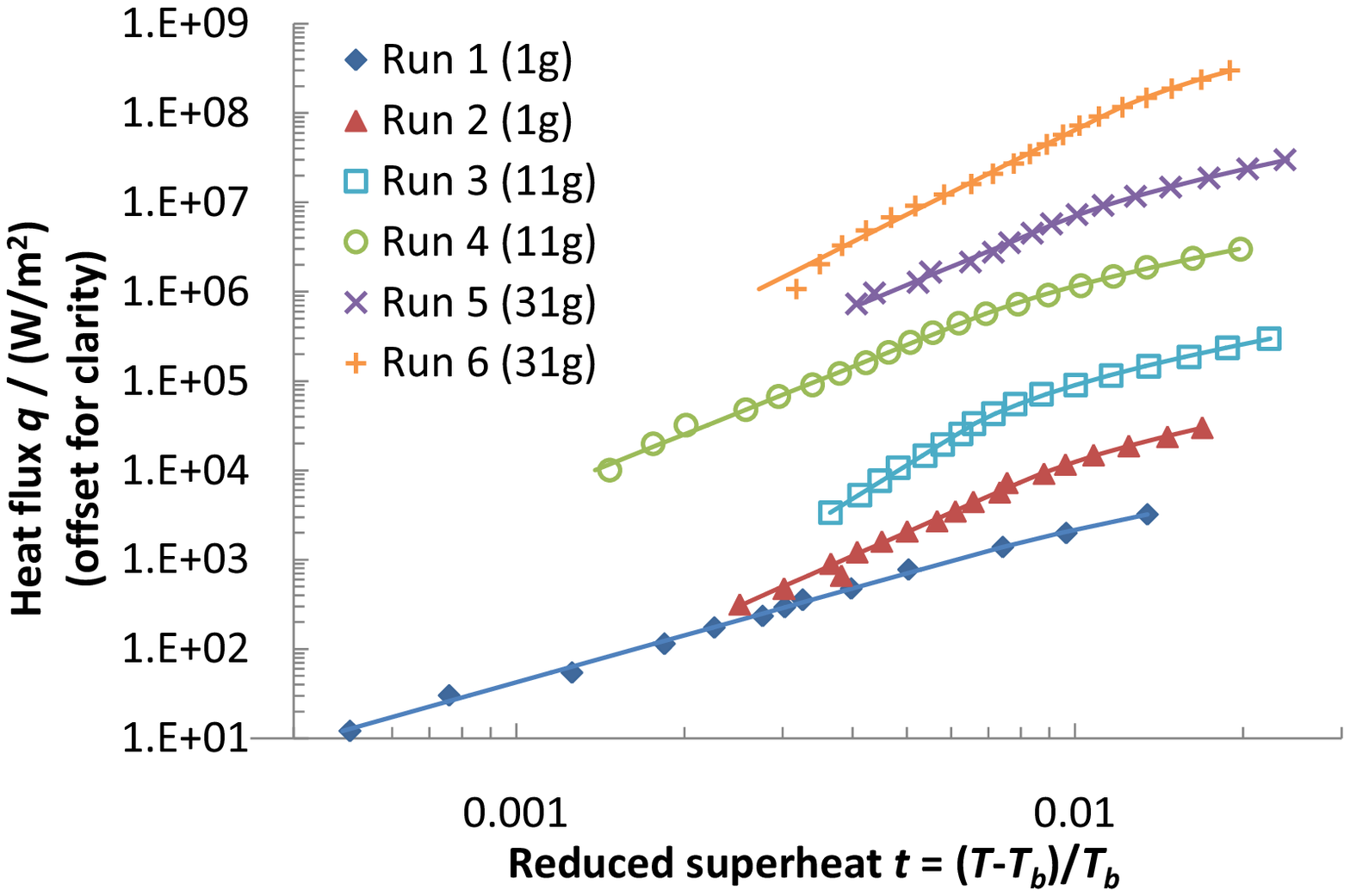}
	\caption{Heat transport data, extracted from time series data similar to those in Fig. \ref{fig:run}, shown on a linear plot (panel a) and the same data on a log-log plot (panel b).  The reduced superheat, calculated from the heater temperature $T$ and the saturation temperature $T_b$, is displayed on the horizontal scale. The vertical scale shows the applied heat flux.  The data have been offset vertically in both plots for legibility( steps of 1000 in panel a and factors of 10 in panel b).  It should be noted that all of the fitting curves pass through $q=0, t=0$ if the offset is removed.  The symbols denote $g_\mathrm{eff}$: closed polygons $g_\mathrm{eff}=1g$, open polygons $g_\mathrm{eff}=6.0g$, crosses $g_\mathrm{eff}=16g$.  The solid lines indicate fits to the data given by Eq. \ref{eqn:fit} and the parameters in Table \ref{table:fits}.}
	\label{fig:fit}
\end{figure}

The reduced superheat we measure is well-fit by a piecewise function of the applied heat flux $q$:
\begin{equation}\label{eqn:fit}
t(q) = \left\{ 
\begin{array}{ll}
a\left(\frac{q}{q_t}\right)^b & \textrm{for }q < q_t ; \\
m\left(\frac{q}{q_t}\right) + c & \textrm{for }q \geq q_t 
\end{array} \right.
\end{equation}
where $q_t$ is some transition heat flux (found by fitting), and the dimensionless quantities $a, b, m,$ and $c$ are fitting parameters subject to smoothness constraints that $t(q)$ and $\partial t(q)/\partial q$.  The constraints are equivalent to
\begin{eqnarray}
m &=& ab\,; \\
c &=& a(1-b),
\end{eqnarray}
leaving three independent free parameters per fit.
Experimental data sets and the resulting fits are shown in Figure \ref{fig:fit} with the fitting parameters listed in Table \ref{table:fits}.  The power law form was chosen for comparison with many empirical and theoretical correlations in the literature\cite{cooper84a, stephan80, rohsenow52, forster59}.  However, this form could not fit the full range of our data.  We have appended the linear portion of the fitting function and smoothness constraints to remedy the poor fits at high heat flux without adding an excessive number of free parameters.  Averages for each value of $g_\mathrm{eff}$ are also shown in Table \ref{table:fits}.  Because of irreproducibility between runs, the uncertainties in the average values are much larger than the uncertainties of individual runs.  The suspect this irreproducibility is caused by uncontrolled transient effects, as discussed below.

\begin{table}[p]
\begin{center}{\scriptsize
\begin{tabular}{ccccccc}
	Run \# & 
	$g_{\mathrm{eff}}/g$ & $q_t/\mathrm{(W\,m^{-2})}$ & $a/10^{-3}$ & $b$ & $m/10^{-3}$ & $c/10^{-3}$ \\
	\hline
	1 &  1 & $1135\pm12$ & $6.58\pm0.05$ & $0.572\pm0.009$ & $3.77\pm0.03$ & $2.81\pm0.06$ \\
	2 &  1 & $729\pm6$ & $7.96\pm0.03$ & $ 0.363\pm0.005$ & $2.89\pm0.02$ & $5.07\pm0.03$ \\
	Average &  1 & $810\pm160$ & $7.59\pm0.66$ & $0.412\pm0.090$ & $3.16\pm0.41$ & $4.62\pm0.89$\\
	\hline
	3  & 6.0 & $272.7\pm0.7$ & $6.27\pm0.01$ & $0.2586\pm0.0008$ & $1.621\pm0.003$ & $4.65\pm0.01$ \\
	4  & 6.0 & $497.4\pm1.1$ & $6.55\pm0.01$ & $0.3983\pm0.0012$ & $2.609\pm0.004$ & $3.94\pm0.01$ \\
	Average  & 6.0 & $340\pm100$ & $6.41\pm0.14$ & $0.302\pm0.064$ & $1.98\pm0.47$& $4.30\pm0.29$ \\
	\hline
	5  & 16 & $586\pm2$ & $9.28\pm0.02$ & $0.386\pm0.001$ & $3.585\pm0.006$ & $5.69\pm0.01$ \\
	6  & 16 & $940\pm4$ & $11.27\pm0.02$ & $0.317\pm0.001$ & $3.574\pm0.009$ & $7.69\pm0.01$\\
	Average & 16 & $660\pm140$ & $10.28\pm0.94$ & $0.352\pm0.029$ & $3.582\pm0.005$ & $6.7\pm1.0$ \\
	\hline
	\multicolumn{2}{c}{Overall average} & $370\pm150$ & $7.2\pm1.6$ & $0.33\pm0.06$
	& $2.3\pm0.8$ & $5.5\pm1.2$
\end{tabular}}
	\caption{Fitting parameters for the data in Fig. \ref{fig:fit}, showing effective gravity $g_\mathrm{eff}$, and parameters from Eq. \ref{eqn:fit}. Average values for each effective gravity and for the entire data set are also shown.}
	\label{table:fits}
	\end{center}
\end{table}

The piecewise boiling curve represented by Eq. \ref{eqn:fit} agrees qualitatively with the results of the centrifuge experiments reported in Refs. \cite{merte90, ulucakli90}, showing different behavior at low and high heat fluxes.  In particular, our low flux data approximately obey the power law $t \propto q^{1/3}$ to within statistical uncertainties; this form has precedent in some of the more well-known correlations, such as Cooper\cite{cooper84a}, Stephan \& Abdelsalam\cite{stephan80}, and Rohsenow\cite{rohsenow52}.  This sublinear behavior may indicate that the number of nucleation sites changes over this range of heat flux.  On possibility is that $q_t$ represents the minimum heat flux required to maintain activity each individual nucleation site.  As the heat flux decreases below this threshold, these sites begin to close off, resulting in decreased heat transfer efficiency.
This exponent does not appear to vary with $g_\mathrm{eff}$, at least to within the uncertainties of our data.

We are unclear about what causes the $t \propto q$ behavior in our high flux data, although the shape of this part of the curve is unmistakable in the linear-scale data plots (see the left panel of Fig. \ref{fig:fit}).  Within the simple picture of nucleation sites discussed the previous paragraph, this is consistent with the conjecture that the number of nucleation sites is independent of heat flux above the threshold flux $q_t$.  In the energy added by increasing heat flux goes only into increasing the superheat, rather than into creating new nucleation sites.

The functional form we have chosen clearly appears in all of our data sets, and the fits for each individual data set are good with the uncertainties in the fitting parameters being a few parts per thousand.  However, because of irreproducibility the averages of runs have large uncertainty.  We have taken some steps to minimize this irreproducibility.

\begin{figure}[p]
	\centering
		\includegraphics[width=0.90\textwidth]{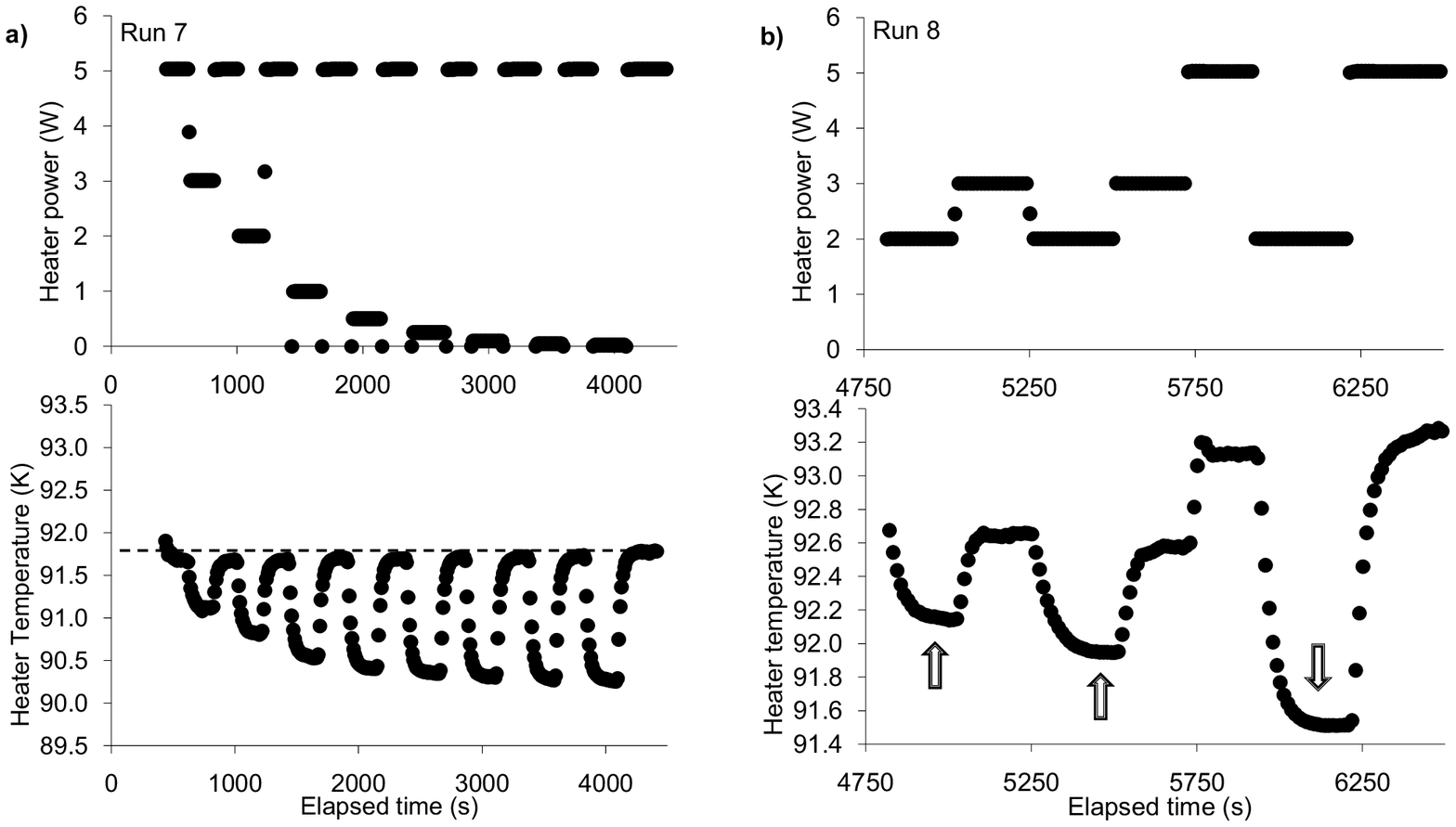}
	\caption{The sequence of heater powers in these data was chosen to demonstrate the memory effect on the heater temperature, as discussed in the text.  In panel a, the power is alternated between a nominal maximum value of $5\unit{W}$ and a sequence of decreasing values.  Note that at each return to $5\unit{W}$, the heater temperature returns to the same value of $91.70\pm0.01\unit{K}$, approximately shown by the dashed line.  In panel b, the power sequence is inverted, starting at a nominal minimum power of $2\unit{W}$ and alternating with an increasing sequence.  In contrast to the data in panel a, these data do not return to the same heater temperature when the heater power returns to $2\unit{W}$, at the areas indicated by the arrows.}
	\label{fig:stairs}
\end{figure}

One technique which improves reproducibility is to avoid starting the system in the metastable natural convective state by starting at a maximum heater power and then decreasing the power.  This sequence minimizes any irreproducibility caused by the hysteretic nature of the transition from natural convection to boiling.  The system seems to retain some memory of the maximum heater power applied over any given experimental run.
To demonstrate this memory effect we produced a pair of runs (Fig. \ref{fig:stairs}) in which the heater power was alternated.  In the first data set, shown in Fig. \ref{fig:stairs}a, the heater power starts at a nominal maximum value, steps to a lower value, returns to the maximum value, steps down to a still lower value, and the pattern repeats.  The key effect in this data is that the heater temperature returns to the same value with each return of the power to the maximum power, indicated by the dashed line in the lower panel of Fig. \ref{fig:stairs}a.  Also, the sequence of lower power steps result in a monotonically decreasing heater temperature sequence.  This pattern is in contrast to what happens if the heater power starts at some low value, increases, returns to the low value, and then increases to a yet higher value.  Such a data set is shown in Fig. \ref{fig:stairs}b.  The important effect in this data is that the temperature at the $2\unit{W}$ steps (indicated by the arrows in the figure) seems to depend on the previous heat value, unlike Fig. \ref{fig:stairs}a. 

We hypothesize that the number of nucleation sites on the heater surface is dependent on the maximum heater power applied since the initiation of boiling.  This could be verified by imaging the heater surface and counting nucleation sites, but our current apparatus lacks this capability.
To ensure the minimum variation between experimental runs, we've chosen a maximum nominal heater power of $5\unit{W}$, corresponding to a maximum heat flux of $3\unit{kW\,m^{-2}}$.  At heater powers higher than $5\unit{W}$ the liquid boils off too quickly for us to acquire sufficient data, exhausting our small sample volume before the system can reach a steady state.

Unfortunately, some irreproducibility between runs persists, most likely because of the transient fluctuations during the preparation phase of the individual data runs (for example, region ``b'' in Fig. \ref{fig:run}).
We believe these oscillations (or occasional lack thereof) cause the remaining variability of the data.  In particular, we observe that the steady-state temperature of the heater at the maximum heater power varies between runs (the rightmost points of the boiling curves in Fig. \ref{fig:fit}).  Unfortunately, in some runs the transient pressure drops below the minimum range of our pressure gauge ($0.5\unit{\textrm{torr}}$), so we are unable to quantify any possible relation between the minimum transient pressure and the heat transport data.  We suspect the variation in maximum temperature between runs is evidence of a variation in the number of nucleation sites.  When the pressure drops during these transient oscillations, the saturation temperature decreases causing a temporary superheated condition in the liquid; the sudden excess of energy in the system may go into activating nucleation sites.  These new nucleation sites may remain active after the system reaches a steady state, resulting in more efficient heat transfer (therefore, lower heater temperature) than would result if the pressure behaved smoothly during the initial stages of boiling.

The irreproducibility caused by the above transient effects may be also obscuring any gravity effect on heat transport.
The values of the fitting parameters given in Table \ref{table:fits} are independent of the applied force to within statistical uncertainties.  This is consistent with previous experiments with gravity values in our range $1 \lesssim g_\mathrm{eff}/g \lesssim 16$  which also find boiling heat transfer to be independent of gravity\cite{merte90, ulucakli90,kim02}.  The large uncertainties in the average fitting parameters are the results of variations among data sets, corresponding to the difficulty in controlling the transient behavior during the setup of the experiment.  A better understanding of the transient pressure effects and/or a larger number of data sets are probably required to resolve any gravity effect in the data.  Comparison of parameter values among the various runs in Table \ref{table:fits} also suggests that some parameters may be correlated, for example, $q_t$ and $a$, although more investigation is needed to confirm any such relation.

\section{Conclusions}
For the saturated nucleate boiling of oxygen, we observe two heat transport behaviors.  At high heat flux values $q\gtrsim 400\unit{W m^{-2}}$, the superheat varies linearly with the applied heat flux.  At lower heat fluxes, the superheat is proportional to $q^{1/3}$.  The latter result is consistent with well established correlations from the literature\cite{cooper84a, stephan80, rohsenow52}.  We are unable to see any change in heat transport behavior caused by varying effective gravity over the range $1g< g_\mathrm{eff} < 16g$, although our analysis is limited by some residual irreproducibility between experimental runs related to transient pressure behavior during the setup phase of the experiment.

Our apparatus has the capability to extend the range of gravity values studied to 
$0 \lesssim g_\mathrm{eff}/g \lesssim 90$.
The larger values of effective gravity can be obtained by increasing the current in our magnet up to its maximum value of $84\unit{A}$.  To study values of $g_\mathrm{eff}$ less than Earth gravity, we can move the sample cell to the bottom end of the solenoid, where the magnetic force is antiparallel to the Earth's gravity (see Fig. \ref{fig:force}b).  For even larger effective gravities or studies of other fluids, such as water, we may also utilize the new Variable Gravity Testbed Facility at the Jet Propulsion Laboratory\cite{chui06}, which is an apparatus similar to ours with a higher maximum magnetic field and a room-temperature bore.

Support for this work was provided by the California Institute of Technology's Jet Propulsion Laboratory, Low Temperature Science and Quantum Sensor Group under contract from the National Aeronautics and Space Administration.  Support for M.E.T. was also provided by the California Institute of Technology Summer Undergraduate Research Fellowships, Tab and Keri Stephens Fellowship. 

\phantomsection
\addcontentsline{toc}{section}{References}
\begin{sloppypar}
\bibliography{lox}
\end{sloppypar}
\end{document}